\documentclass[%
 showpacs,
 reprint,
 amsmath,amssymb,
 aps,
]{revtex4-1}

\usepackage{graphicx}
\usepackage{dcolumn}
\usepackage{bm}
\usepackage[colorlinks,linkcolor=blue, urlcolor=blue, anchorcolor=blue, citecolor=blue]{hyperref}
\usepackage{multirow}
\usepackage{indentfirst}
\usepackage{amsthm}
\usepackage{amsmath}
\usepackage{epstopdf}
\usepackage{array}
\usepackage{float}
\usepackage[marginal]{footmisc}

\begin{document}

\preprint{APS/123-QED}

\title{Discrimination Between Quantum Common Causes and Quantum Causality}
\author{Mingdi Hu}
\email{mingdihu@tju.edu.cn}
\affiliation{School of Computer Science and Technology, Tianjin University, No. 135, Ya Guan Road,
 Tianjin, China}
\author{Yuexian Hou}
\email{corresponding author: yxhou@tju.edu.cn}
\affiliation{School of Computer Science and Technology, Tianjin University, No. 135, Ya Guan Road,
 Tianjin, China}

\begin{abstract}
In classic cases, Reichenbach's principle implies that discriminating between common causes and causality is unprincipled since the discriminative results essentially depend on the selection of possible conditional variables. For some typical quantum cases, K.Reid $et$ $al$. \href{https://www.nature.com/articles/nphys3266}{[Nat. Phys. 11, 414 (2015)]} presented the statistic $C$ which can effectively discriminate quantum common causes and quantum causality over two quantum random variables (i.e., qubits) and which only uses measurement information about these two variables. In this paper, we formalize general quantum common causes and general quantum causality. Based on the formal representation, we further investigate their decidability via the statistic $C$ in general quantum cases. We demonstrate that (i) $C \in \left[ { - 1,\frac{1}{{27}}} \right]$ if two qubits are influenced by quantum common causes; (ii) $C \in \left[ { - \frac{1}{{27}},1} \right]$ if the relation between two qubits is quantum causality; (iii) a geometric picture can illuminate the geometric interpretation of the probabilistic mixture of quantum common causes and quantum causality. This geometric picture also provides a basic heuristic to develop more complete methods for discriminating the cases corresponding to $C \in \left[ { - \frac{1}{{27}},\frac{1}{{27}}} \right]$. Our results demonstrate that quantum common causes and quantum causality can be discriminated in a considerable scope.
\end{abstract}

\pacs{03.67.-a}

\maketitle

\section{Introduction}
\begin{figure}[!htp]
 \centering
 \includegraphics[width=2.0in,height=1.2in]{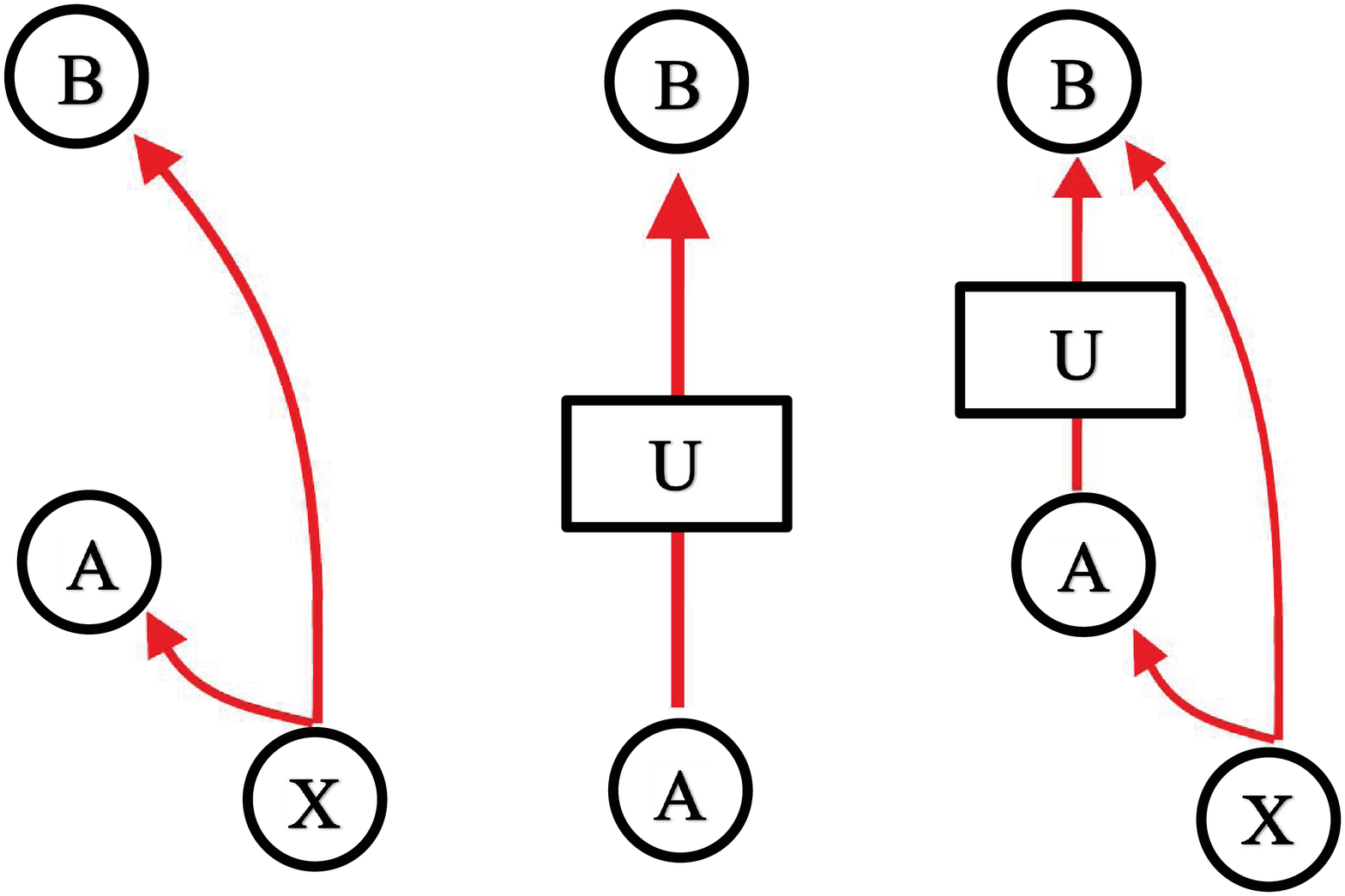}
 \caption{\label{figure_1} Three possible relations of two qubits $A, B$ \cite{Pearl2000Causality,Spirtes2000Causation}. Left to right: quantum common causes (common causes $X$ influence on $A$ and $B$), quantum causality ($B = {\bf{U}} \cdot A$), and a mixture of both, where nodes represent qubits, directed edges represent causal influences, and ${\bf{U}}$ represents the quantum causality transformation.}
 \end{figure}

It is a scientific problem to discriminate common causes and causality. The principle of causal explanation was first put forward explicitly by Reichenbach \citep{Reichenbach}: if two physical variables $A$ and $B$ are statistically correlated (to be exact, they are dependent), then they can be explained as follows: (i)common causes, which mean that there are common causes influencing both $A$ and $B$; (ii)causality, namely, direct cause, which means $A$ $(B)$ directly causes $B$ $(A)$. To some extent, the above two definitions are informal and non-operational. Therefore, a central problem is how to discriminate them by means of data.\\
\indent In classic cases, Reichenbach's principle \citep{Reichenbach} suggestes that only if $p\left( {A,B} \right) \ne p\left( A \right)p\left( B \right)$ and $p\left( {A,B|X} \right) \ne p\left( {A|X} \right)p\left( {B|X} \right)$ hold, where $\{ X\}$ represents the family of all possible common cause sets, is it reasonable to infer that there exists a causality between $A$ and $B$. However, it is often difficult to determine the set of all possible conditional variables. Even if $\{ X\}$ can be properly defined, it often requires a large number of samples to compute the statistics on $\{ X\}$. Consequently, the discrimination between common causes and causality is difficult and heavily dependent on prior knowledge. Hence the motto ``Correlation does not imply causation'' was coined.\\
\indent In quantum cases, quantum common causes and quantum causality (also known as quantum direct cause) can be formally defined. And hence they can be exactly discriminated, at least, in a considerable scope. Actually, quantum causal inference does not depend on a conditional variables family $\{ X\}$, but only uses the measurement information on the two quantum random variables considered (i.e., qubits).\\
\indent Quantum correlation research dates back to at least Bell, who \citep{Bell1964On} pioneered the study of non-classical characters of Einstein-Podolsky-Rosen-like correlations. Subsequently, great progresses was made in the study of spatial correlations \citep{Henderson2001Classical,Hellweg2003Measurement,Ollivier2001Quantum,Bobrov2014Imaging,Daki2010Necessary} and time correlations \citep{Karsch2012Signatures,Sergi2015Time,Pawlowski2015Real,Fitzsimons2015Quantum,Allen2017Quantum}. For time correlations, Fitzsimons $et$ $al$. \citep{Fitzsimons2015Quantum} defined a pseudo-density operator of a temporally ordered bipartite quantum system. Additionally, they demonstrated that an irregular pseudo-density operator implied that there existed quantum causality between two qubits. Reid $et$ $al$. \cite{Ried2015A} developed the work of Fitzsimons $et$ $al$., and presented a real statistic $C$ to experimentally assess the existence of causality in some typical cases [i.e., (i) a possible quantum common cause can be represented as one of four Bell states; (ii) a possible quantum causality can be represented as one of four Pauli matrices].\\
\indent Motivated by their work, this paper focuses on the discriminant between quantum common causes and quantum causality in general cases. To this end, first, we generalize the formal representation of the relation of two quantum random variables (i.e., qubits): (i)any four-dimensional density operator $\rho $ corresponds to a possible quantum common cause and vice versa; here, the quantum common causes include not only the usual quantum correlations (i.e., the non-canonical correlation that is induced by entanglement) but also product states and their mixtures, for example, $\rho =\frac{1}{2}\left| {{00}} \right\rangle \left\langle {{00}} \right| + \frac{1}{2}\left| {{11}} \right\rangle \left\langle {{11}} \right|$; (ii)any element in ${\bf{U}}(2)$ corresponds to a possible quantum causality and vice versa; (iii) there can be a mixture of the above two cases, as shown in Fig. \ref{figure_1}.\\

\indent Based on above representation, this paper theoretically demonstrates the bound of the statistic $C$ in cases (i) and (ii), the results are shown in Sec. \ref{sec:II} and \ref{sec:III}, respectively. In Sec. \ref{sec:IV}, a geometric picture is presented to illuminate the geometric interpretation of case (iii). In Sec. \ref{sec:V}, a method is proposed to distinguish the overlapped area of cases (i) and (ii). In Sec. \ref{sec:VI}, we summarize and propose the future work.

\section{QUANTUM COMMON CAUSES}\label{sec:II}
In this section, we first review the statistic $C$. Reid $et$ $al$. \cite{Ried2015A} presented a scalar statistic $C \equiv \prod\limits_{i = 1}^3 {{C_{ii}}}$ to indicate the following two illuminating cases: (i) if $C=+1$, then it indicates the quantum causality transformations (quantum direct cause) corresponding to four Pauli operators, i.e., $\sigma_{i}, i=0, \ldots, 3$; and (ii) if $C =-1$, then it indicates quantum common causes entailing perfect correlations or anticorrelations when measured by Pauli observables. In this case, they are four Bell states. Table \ref{table:1} displays more details. Apparently, the statistic $C$ can only take a value 1 or -1, which limits the discriminant between quantum common causes and quantum causality. Therefore, it is necessary to extend the scalar $C$ to the continuous real domain.\\

\begin{table}[!htp]
 \scalebox{1.0}{
 \begin{tabular}{|c|c|c|c|c|c|}
 \hline
 \multicolumn{4}{|c|}{Pattern of}&Causality&Common\\
 \multicolumn{4}{|c|}{correlations}&(Direct cause) &cause \\
 \hline
 ${{C_{11}}}$&${{C_{22}}}$&${{C_{33}}}$&$C \equiv \prod\limits_{i = 1}^3 {{C_{ii}}}$& & \\
 \hline
 +1&+1&+1&+1&${\bf{U}}={\sigma _0}$&No\\
 +1&-1&-1&+1&${\bf{U}}={\sigma _1}$&No\\
 -1&+1&-1&+1&${\bf{U}}={\sigma _2}$&No\\
 -1&-1&+1&+1&${\bf{U}}={\sigma _3}$&No\\
 \hline
 +1&-1&+1&-1&No&$\rho =\left| {{b_1}} \right\rangle \left\langle {{b_1}} \right|$\\
 -1&+1&+1&-1&No&$\rho =\left| {{b_2}} \right\rangle \left\langle {{b_2}} \right|$\\
 +1&+1&-1&-1&No&$\rho =\left| {{b_3}} \right\rangle \left\langle {{b_3}} \right|$\\
 -1&-1&-1&-1&No&$\rho =\left| {{b_4}} \right\rangle \left\langle {{b_4}} \right|$\\
 \hline
 \end{tabular}}
 \caption{\label{table:1} Signatures of causal structure \cite{Ried2015A}. Assume that the same Pauli observable ${\sigma _i}$ is measured on two qubits, i.e., $\left( {i,i} \right) \in \left\{ {\left( {1,1} \right),\left( {2,2} \right),\left( {3,3} \right)} \right\}$, outcomes are $k$ and $m$. Correlation indices ${C_{ii}} \equiv  p\left( {k = m|ii} \right) - p\left( {k \ne m|ii} \right) \left( {i = 1,2,3} \right)$. A possible quantum causality is one of four Pauli matrices ${{\bf{\sigma}} _0} = {\bf{I}} = \left( {\begin{array}{*{20}{c}}
1&0\\
0&1
\end{array}} \right), {{\bf{\sigma}} _1} = {\bf{X}} = \left( {\begin{array}{*{20}{c}}
0&1\\
1&0
\end{array}} \right), {{\bf{\sigma}} _2} = {\bf{Y}} = \left( {\begin{array}{*{20}{c}}
0&{ - i}\\
i&0
\end{array}} \right), {{\bf{\sigma}} _3} = {\bf{Z}} = \left( {\begin{array}{*{20}{c}}
1&0\\
0&{ - 1}
\end{array}} \right)$. A possible quantum common cause is one of four Bell states: ${{\rm{|}}{{\rm{b}}_{\rm{1}}}\rangle } = \frac{{|00 >  + |11 > }}{{\sqrt 2 }}{\kern 1pt} {\kern 1pt}, {{\rm{|}}{{\rm{b}}_{\rm{2}}}\rangle } = \frac{{|00 >  - |11 > }}{{\sqrt 2 }}, {{\rm{|}}{{\rm{b}}_{\rm{3}}}\rangle } = \frac{{|01 >  + |10 > }}{{\sqrt 2 }}{\kern 1pt} {\kern 1pt}, {{\rm{|}}{{\rm{b}}_{\rm{4}}}\rangle } = \frac{{|01 >  - |10 > }}{{\sqrt 2 }}$.}
\end{table}

In the following, we show that ${C_{\rm{CC}}} \in [ - 1,\frac{1}{{27}}]$ in the cases of the general quantum common causes, where ${C_{\rm{CC}}}$  means the statistic $C$ in the cases of quantum common causes. The general quantum common causes can be any four-dimensional density operator $\rho$ and vice versa; here, the quantum common causes include not only the usual quantum correlations (i.e., the non-canonical correlation that is induced by entanglement) but also the possible correlations induced by the mixture of product states. Additionally, we demonstrate that ${C_{\rm{CC}}} \in [ - 1,\frac{1}{{27}}]$ holds too in the cases of quantum correlations.\\
\indent Now, we analyze the bound of ${C_{\rm{CC}}}$ (i.e., $C$ in cases of quantum common causes) in detail. If the same Pauli observable $\sigma_{i} \left( {i = 1,2,3} \right)$ is measured on the two qubits, outcomes are $k$ and $m$ respectively, then
 \begin{equation}
 \begin{split}
{C}({\rho } ) &= \prod\limits_{i = 1}^3 {{C_{ii}}} (\rho ) \\
& = \prod\limits_{i = 1}^3 [ p\left( {k = m|ii} \right) - p\left( {k \ne m|ii} \right)] \in [- 1,1].
 \end{split}\label{Eq:C(rho)}
 \end{equation}
It is easy to check that ${C_{\rm{CC}}}({\rm{\rho }} )= -1$ if the two-qubit state is one of four Bell states. Therefore, we only need to solve the supremum bound of ${C_{\rm{CC}}}(\rho )$. To this end, we need the following definition:\\
 \indent \emph{Definition 1.} The vector-valued function ${{\bf{P}}}{\bf{(\rho )}}$ on the density operator $\rho $ is defined as
 \begin{equation}
 {{\bf{P}}}{\bf{(\rho )}} \equiv \left( {\begin{array}{*{20}{c}}
{{C_{11}}(\rho )}\\
{{C_{22}}(\rho )}\\
{{C_{33}}(\rho )}
\end{array}} \right)\label{def:C(rho)}.
\end{equation}
\indent When $\rho $ is a pure state, it is equivalent to define ${\bf{P}}(|{\rm{\varphi }}\rangle)$ on a state vector $|{\rm{\varphi }}\rangle$. Then lemma 1 is obtained.\\

\indent {\bf{\emph{Lemma 1.}}} For all $ |{\rm{\varphi }}\rangle\in{\mathbb{R}^4}$, ${{\bf{P}}}(|{\rm{\varphi }}\rangle )$ forms a regular tetrahedron ${\bf{T}}_{\rm{CC}}$ with vertices ${{\bf{P}}}\left( {|{b_1}\rangle } \right) = (1 ,{ - 1} ,  1)^{'}$, ${{\bf{P}}}\left( {|{b_2}\rangle } \right) = ({ -1 },1,1)^{'}$, ${{\bf{P}}}\left( {|{b_3}\rangle } \right) = (1 ,1 ,{ - 1})^{'}$, and ${{\bf{P}}}\left( {|{b_4}\rangle } \right) = ( -1 , -1 ,{ - 1})^{'}$.\\
\begin{proof}
\indent Four Bell states $|{b_j}\rangle  \in{\mathbb{R}^4}$$\left( {j = 1, \ldots ,4} \right)$, and ${\bf{P}}\left( {|{b_j}\rangle } \right)$$\left( {j = 1, \ldots ,4} \right)$ form four vertices of a regular tetrahedron in $\mathbb{R}^4$. Clearly, four Bell states are a set of standard orthonormal basis in $\mathbb{R}^4$. So any pure state $|{\rm{\varphi }}\rangle $ in $\mathbb{R}^4$ can be represented as
\begin{equation}
|{\rm{\varphi }}\rangle  = \sum\limits_{j = 1}^4 {{{\rm{w}}_{{j}}}  {\rm{|}}{{\rm{b}}_{{j}}}\rangle ,}\label{Eq:phi}
\end{equation}
where ${w_j} \in \mathbb{R}\left( {j = 1, \ldots ,4} \right)$ and $\mathop \sum \limits_{{{j}} = 1}^4 {{\rm{w}}_{{j}}}^2 = 1$.\\
\indent Then, according to Eq. (\ref{Eq:phi}), Eq. (\ref{Eq:C(phi)}) is obtained (see Supplemental Material for proof \cite{Authors12}).
\begin{equation}
{{\bf{P}}}(|{\rm{\varphi }}\rangle ) = \mathop \sum \limits_{{{j}} = 1}^4 {{{\rm{w}}_{{j}}}^2}{{\bf{P}}}\left( {|{b_j}\rangle } \right) .\label{Eq:C(phi)}
\end{equation}
\indent Hence, $\forall |{\rm{\varphi }}\rangle\in{\mathbb{R}^4}$, ${{\bf{P}}}(|{\rm{\varphi }}\rangle ) \in {{{\bf{T}}_{\rm{CC}}}}$, where ${\bf{T}}_{\rm{CC}}$ is a regular tetrahedron with four vertices ${{\bf{P}}}\left( {|{b_j}\rangle } \right)\left( {j = 1, \ldots ,4} \right)$, as shown in Fig. \ref{figure_2}.\\
\indent On the other hand, for a point in ${\bf{T}}_{\rm{CC}}$, this point can be represented as $\mathop \sum \limits_{{{j}} = 1}^4 {{\rm{w}}_{{j}}}^2{{\bf{P}}}\left( {|{b_j}\rangle } \right)$, where ${w_j} \in \mathbb{R}\left( {j = 1, \ldots ,4} \right)$ and $\mathop \sum \limits_{{{j}} = 1}^4 {{\rm{w}}_{{j}}}^2 = 1$. The pure quantum state $|{\rm{\varphi }}\rangle $ corresponding to this point can be represented as
\begin{equation}
|{\rm{\varphi }}\rangle  = \sum\limits_{j = 1}^4 {{{\rm{w}}_{{j}}}  {\rm{|}}{{\rm{b}}_{{j}}}\rangle } .\label{Eq:wj}
\end{equation}
\end{proof}
\indent {\bf{\emph{Lemma 2}.}} For all $ |{\rm{\varphi }}\rangle\in{\mathbb{R}^4}$, ${C_{\rm{CC}}}(|{\rm{\varphi }}\rangle ) \le \frac{1}{{27}}$.
\begin{proof}
\indent Maximizing ${C_{\rm{CC}}}(|{\rm{\varphi }}\rangle ) = \prod\limits_{i = 1}^3 {{C_{ii}}(|{\rm{\varphi }}\rangle )} $ under the condition of ${{\bf{P}}}(|{\rm{\varphi }}\rangle ) \in {{{\bf{T}}_{\rm{CC}}}}$ is equivalent to Eq. (\ref{Eq:kkt1}).
\begin{equation}
\begin{array}{l}
\mathop {\max }\limits_{{\kern 1pt} |{\rm{\varphi }}\rangle  \in {\mathbb{R}^4}} {\kern 1pt} {\kern 1pt} {C_{\rm{CC}}}(|{\rm{\varphi }}\rangle )\\
s.t.,{\kern 1pt} {\kern 1pt} {\kern 1pt} {\kern 1pt} \langle {\rm{\varphi }}{\kern 1pt} |{\rm{\varphi }}\rangle  = 1.
\end{array}\label{Eq:kkt1}
\end{equation}
\indent According to Eq. (\ref{Eq:kkt1}), apparently, the feasible region is a convex set. And the objective function ${C_{\rm{CC}}}(|{\rm{\varphi }}\rangle )$ is a simple cubic function, although it is not a convex function; it is convenient to construct a Lagrangian function $F(|{\rm{\varphi }}\rangle ,\lambda )$ to solve all extreme points.

\begin{equation}
F(|{\rm{\varphi }}\rangle ,\lambda ) = {C_{\rm{CC}}}(|{\rm{\varphi }}\rangle ) - \lambda (\langle {\rm{\varphi }}{\kern 1pt} |{\rm{\varphi }}\rangle  - 1) ,\label{Eq:F}
\end{equation}
\indent All local extreme points (134 in total, see Supplemental Material \cite{Authors12}) are solved with Karush $-$ Kuhn $-$ Tucker (KKT) conditions. And the maximum
\begin{equation}
\max {C_{\rm{CC}}}(|{\rm{\varphi }}\rangle ) = \frac{1}{{27}} .
\end{equation}
\end{proof}
\indent {\bf{\emph{Lemma 3}.}}For all $ |\phi \rangle  \in {\mathbb{C}^4}$, ${C_{\rm{CC}}}(|\phi \rangle ) \le \frac{1}{{27}}$.
\begin{proof}
\indent Given an arbitrary pure quantum state $|\phi \rangle  = \left( {\begin{array}{*{20}{c}}
{a + bi}\\
{c + di}\\
{m + ni}\\
{p + qi}
\end{array}} \right)$,$a,b,c,d,m,n,p,q \in \mathbb{R}$, it can be decomposed into Eq. (\ref{Eq:phi2}).\\
\begin{equation}
|\phi \rangle  = cos{\kern 1pt} \alpha   \left| {x\rangle  + {\rm{ }}sin{\kern 1pt} \alpha   } \right|y\rangle   i .\label{Eq:phi2}
\end{equation}

\indent where $cos{\kern 1pt} \alpha = \pm \sqrt {{a^2} + {c^2} + {m^2} + {p^2}}$, $sin{\kern 1pt} \alpha =\pm \sqrt {{b^2} + {d^2} + {n^2} + {q^2}}$, and $|x\rangle  = \frac{1}{\cos {\kern 1pt} \alpha }  \left( {\begin{array}{*{20}{c}}
{{a}}\\
{\begin{array}{*{20}{c}}
{{c}}\\
{{m}}
\end{array}}\\
{{p}}
\end{array}} \right)$, $|y\rangle  = \frac{1}{\sin {\kern 1pt} \alpha }  \left( {\begin{array}{*{20}{c}}
{\rm{b}}\\
{\begin{array}{*{20}{c}}
{\rm{d}}\\
{\rm{n}}
\end{array}}\\
{\rm{q}}
\end{array}} \right)$ when $cos {\kern 1pt}\alpha \ne 0$ and $sin {\kern 1pt}\alpha \ne 0$. Specially, when $cos{\kern 1pt} \alpha = 0$ or $sin{\kern 1pt} \alpha = 0$, $|x\rangle  = {\mathbf{0}}$ or $|y\rangle  = {\mathbf{0}}$ (i.e., $|\phi \rangle  = |y\rangle  i $ or $|\phi \rangle  = |x\rangle $).\\
\indent Then, according to Eq. (\ref{Eq:phi2}), Eq. (\ref{Eq:C(psi)}) is further calculated and obtained (see Supplemental Material for proof \cite{Authors12})
\begin{equation}
{\bf{P}}(|\phi \rangle ) = co{s^2}\alpha   {\bf{P}}(|x\rangle ) + {\rm{ }}si{n^2}\alpha   {\bf{P}}(|y\rangle ).\label{Eq:C(psi)}
\end{equation}
\indent According to Eq. (\ref{Eq:C(psi)}) and lemma 1, for $\forall |\phi \rangle  \in {\mathbb{C}^4}$, there must exist $|{\rm{\varphi }}\rangle  \in {\mathbb{R}^4}$ such that ${{\bf{P}}}(|\phi \rangle ) = {{\bf{P}}}(|{\rm{\varphi }}\rangle )$ holds. According to lemma 2, ${C_{\rm{CC}}}(|\phi \rangle ) \le \frac{1}{{27}}$.
\end{proof}
\indent {\bf{\emph{Theorem 1}.}} $\rho $ is an arbitrary density operator of a 2-qubit system, ${C_{\rm{CC}}}(\rho ) \in [ - 1,\frac{1}{{27}}]$.
\begin{proof}
\indent Lemma 3 has proved that ${C_{\rm{CC}}}(\rho ) \le \frac{1}{{27}}$ when $\rho$ is a pure state. When $\rho$ is a mixed state, it can be regarded as a convex combination of several pure states. According to the lemmas 1 and 2, there must exist $|{\rm{\varphi }}\rangle  \in {\mathbb{R}^4}$ such that ${C_{\rm{CC}}}(\rho ) = {C_{\rm{CC}}}(|{\rm{\varphi }}\rangle )$ holds. Thus, ${C_{\rm{CC}}}(\rho ) \in [ - 1,\frac{1}{{27}}]$.
\end{proof}
\indent In general quantum common causes, ${C_{\rm{CC}}} \in [ - 1,\frac{1}{{27}}]$ is proved in Theorem 1. Note that the lower bound (-1) and the upper bound ($\frac{1}{27}$) are also tight in terms of the quantum entanglement states since the lower bound (-1) and the upper bound ($\frac{1}{27}$) can be by approached by, e.g., Bell states or $|{\rm{\varphi }}\rangle = \left( {\begin{array}{*{20}{c}}
{ - \frac{2}{{\sqrt 6 }}}\\
{\frac{1}{{\sqrt 6 }}}\\
{ - \frac{1}{{\sqrt 6 }}}\\
0
\end{array}} \right)$, respectively.

\section{QUANTUM CAUSALITY}\label{sec:III}
 What is quantum causality? Given a single-qubit system $A$, $A$ is measured. After a unitary evolution, $A$ becomes a new single-qubit system $B$. $B$ is measured again. Then quantum causality means that the measurement result of $B$ is causally influenced by a certain unitary evolution on the measurement result of $A$. \\
 \indent In the following, we show ${C_{\rm{DC}}} \in [ - \frac{1}{{27}},1]$ in the cases of the general quantum causality, where ${C_{\rm{DC}}}$  means the statistic $C$ in cases of quantum direct causes (also known as quantum causality). The general quantum causality can be each element in ${\bf{U}}(2)$ and vice versa.\\
 \indent Now, we analyze the bound of ${C_{\rm{DC}}}$ (i.e., $C$ in cases of quantum direct cause) in detail. According to Eq. $(\ref{Eq:C(rho)})$, ${C_{\rm{DC}}} \le 1$. And it is easy to check ${C_{\rm{DC}}}=1$ when ${\bf{U}}= \sigma_{i}\left({i = 0, \ldots ,3} \right)$. So we just need to prove the infimum of ${C_{\rm{DC}}}$. First, we observe lemma 4 and its proof.

\indent {\bf{\emph{Lemma 4}.}} ${C_{\rm{DC}}}$ only depends on $\bf{U}$; it is invariant to $\rho$, where $\rho$ is the initial state of a single-qubit system.

 \begin{proof}
 Suppose that $\rho $ is a pure state of the single-qubit system $A$, here, $\rho \in {\mathbb{C}}^{2 \times 2}$. $A$, $B$ are measured respectively by the Pauli matrix ${\bf{X}}$. Two measurement results include two cases: (i) Both $A$ and $B$ are collapsed to $|{{{{x}}_0}} \rangle$ or $|{{{{x}}_1}} \rangle$ (two eigenstates of ${\bf{X}}$); (ii) $B$ is collapsed to $|{{{{x}}_1}} \rangle$ or $|{{{{x}}_0}} \rangle$ under the condition that $A$ is collapsed to $|{{{{x}}_0}} \rangle$ or $|{{{{x}}_1}} \rangle$. According to Eq. (\ref{Eq:C(rho)}), then

\begin{equation}
\begin{split}
{C_{11}} &= {p_{\rm{A}}}\left( {{{|{{{{x}}_0}} \rangle}}} \right)  {p_{{{B}}|{{A}}}}\left( {{{|{{{{x}}_0}} \rangle}}} \right) + {p_{\rm{A}}}\left( {{{|{{{{x}}_1}} \rangle}}} \right)  {p_{{{B}}|{{A}}}}\left( {{{|{{{{x}}_1}} \rangle}}} \right) \\
&-\left\{ {1 - [{p_{\rm{A}}}\left( {{{|{{{{x}}_0}} \rangle}}} \right)  {p_{{{B}}|{{A}}}}\left( {{{|{{{{x}}_0}} \rangle}}} \right) + {p_{\rm{A}}}\left( {{{|{{{{x}}_1}} \rangle}}} \right)  {p_{{{B}}|{{A}}}}\left( {{{|{{{{x}}_1}} \rangle}}} \right)]} \right\},
\end{split}\label{Eq:C11}
\end{equation}
where ${p_{\rm{A}}}\left( {{{|{{{{x}}_0}} \rangle}}} \right)$ or ${p_{\rm{A}}}\left( {{{|{{{{x}}_1}} \rangle}}} \right)$ is the probabilities that $A$ is collapsed to $|{{{{x}}_0}} \rangle$ or $|{{{{x}}_1}} \rangle$. ${p_{{{B}}|{{A}}}}\left( {{{|{{{{x}}_0}} \rangle}}} \right)$ or ${p_{{{B}}|{{A}}}}\left( {{{|{{{{x}}_1}} \rangle}}} \right)$ is the conditional probability that $B$ is collapsed to $|{{{{x}}_0}} \rangle$ or $|{{{{x}}_1}} \rangle$ under the condition that $A$ is collapsed to $|{{{{x}}_0}} \rangle$ or $|{{{{x}}_1}} \rangle$. And
 \begin{equation}
 \begin{split}
 {p_{{{B}}|{{A}}}}\left( |{{x_0}}\rangle \right) = {\left( {{\bf{U}}  |{{x_0}}\rangle } \right)^{'}}  {{\bf{P}}_{x_0}}  \left( {{\bf{U}}  |{{x_0}}\rangle } \right),\\
  {p_{{{B}}|{{A}}}}\left( |{{x_1}}\rangle \right) = {\left( {{\bf{U}}  |{{x_1}}\rangle } \right)^{'}}  {{\bf{P}}_{x_1}}  \left( {{\bf{U}}  |{{x_1}}\rangle } \right),
  \end{split}\label{p(B|A)}
 \end{equation}

 where ${\bf{U}}$ is the causal evolution; ${{\bf{P}}_{x_0}} = |{{x_0}}\rangle \left\langle {{{x_0}}|} \right.$ and ${{\bf{P}}_{x_1}} = |{{x_1}}\rangle \left\langle {{{x_1}}|} \right.$ are measurement operators.\\
 \indent We aim to prove ${C_{\rm{DC}}}$ is invariant to $\rho$; according to Eq. (\ref{Eq:C(rho)}), we just need to prove ${C_{ii}} \left( {i = 1, \ldots ,3} \right)$ is invariant to $\rho$. First, we prove Eq. (\ref{Eq:C11}) is invariant to $\rho$.

 \indent According to Eq. (\ref{p(B|A)}), it is convenient to prove ${p_{{{B}}|{{A}}}}\left( |{{x_0}}\rangle \right)={p_{{{B}}|{{A}}}}\left( |{{x_1}}\rangle \right)$. Because ${\bf{U}}  |{{x_0}}\rangle $(${\bf{U}}  |{{x_1}}\rangle $) means a rotation of $|{{x_0}}\rangle $($|{{x_1}}\rangle $), and the angle of $|{{x_0}}\rangle$ and $|{{x_1}}\rangle$ is same as the angle of ${\bf{U}}  |{{x_0}}\rangle $ and ${\bf{U}}  |{{x_1}}\rangle $. Therefore, there must exist
\begin{equation}
{p_{{{B}}|{{A}}}}\left( |{{x_0}}\rangle \right) = {p_{{{B}}|{{A}}}}\left( |{{x_1}}\rangle \right) .\label{p(B|A(x0=x1))}
\end{equation}
\indent Clearly, ${p_{\rm{A}}}\left( {{{|{{{{x}}_0}} \rangle}}} \right) + {p_{\rm{A}}}\left( {{{|{{{{x}}_1}} \rangle}}} \right) =1$, according to Eq. (\ref{p(B|A(x0=x1))}), and Eq. (\ref{Eq:C11}) is simplified as 
\begin{equation}
C_{11}=2{p_{{{B}}|{{A}}}}\left( |{{x_0}}\rangle \right)-1=2{p_{{{B}}|{{A}}}}\left( |{{x_1}}\rangle \right)-1. \label{C_11}
\end{equation}
\indent Similarly, ${C_{22}} = 2{p_{{{B}}|{{A}}}}\left( |{{y_0}}\rangle \right) - 1$, and ${C_{33}} = 2{p_{{{B}}|{{A}}}}\left( |{{z_0}}\rangle \right) - 1$. According to Eq. (\ref{p(B|A)}), ${p_{{{B}}|{{A}}}}\left( |{{x_0}}\rangle \right)$ only depends on ${\bf{U}}$, it is invariant to $\rho$. Therefore, ${C_{ii}}\left( {i = 1, \ldots ,3} \right)$ only depend on ${\bf{U}}$, they are invariant to $\rho$.\\
\indent The above conclusion is easy to extend to the case of mixed states. When ${\bf{U}}$ is given, $\rho $ is a mixture of several pure states $|{\rm{\varphi }}\rangle $, and each pure state corresponds to the same ${C_{ii}}\left( {i = 1, \ldots ,3} \right)$. Thus, ${C_{ii}}(\rho ){\rm{ = }}{C_{ii}}(|\varphi \rangle )$.\\
\end{proof}

\indent According to lemma 4, formally, we define ${{\bf{P}}}\left( {\bf{U}} \right)$ as follows.\\
\indent \emph{Definition 2.} The vector-valued function ${{\bf{P}}}\left( {\bf{U}} \right)$ on the unitary matrix ${\bf{U}}$ is defined as
\begin{equation}
{{\bf{P}}}\left( {\bf{U}} \right) \equiv \left( {\begin{array}{*{20}{c}}
{{C_{11}}\left( {\bf{U}} \right)}\\
{{C_{22}}\left( {\bf{U}} \right)}\\
{{C_{33}}\left( {\bf{U}} \right)}
\end{array}} \right) \label{def:C(U)}.
\end{equation}
\indent {\bf{\emph{Lemma 5}.}} For all $ {\bf{U}} \in {\bf{U}}\left( 2 \right)$, there must exist ${p_{{j}}} \ge 0\left( {j = 0, \ldots ,3} \right)$, $\mathop \sum \limits_{{{j}} = 0}^3 {p_{{j}}} = 1$ such that ${{\bf{P}}}\left( {\bf{U}} \right) = \mathop \sum \limits_{{{j}} = 0}^3 {p_{{j}}}  {{\bf{P}}}\left( {{\sigma _{{j}}}} \right)$ holds, where ${\mathbf{P}}{\text{(}}{\sigma _0}) = \left( {\begin{array}{*{20}{c}}
  1 \\
  1 \\
  1
\end{array}} \right)$,  ${\mathbf{P}}{\text{(}}{\sigma _1}) = \left( {\begin{array}{*{20}{c}}
  1 \\
  -1 \\
  -1
\end{array}} \right)$,  ${\mathbf{P}}{\text{(}}{\sigma _2}) = \left( {\begin{array}{*{20}{c}}
  -1 \\
  1 \\
  -1
\end{array}} \right)$,  ${\mathbf{P}}{\text{(}}{\sigma _3}) = \left( {\begin{array}{*{20}{c}}
  -1 \\
  -1 \\
  1
\end{array}} \right)$.

\begin{proof}
An arbitrary unitary matrix ${\bf{U}} \in {\bf{U}}\left( 2 \right)$ can be parameterized as
\begin{equation}
{\bf{U}} = \left( {\begin{array}{*{20}{c}}
{{a_1} + {a_2}i}&{{b_1} + {b_2}i}\\
{ - {e^{\alpha {\rm{i}}}}\left( {{b_1} - {b_2}i} \right)}&{{e^{\alpha {\rm{i}}}}\left( {{a_1} - {a_2}i} \right)}
\end{array}}  \right) ,\label{Eq:U}
\end{equation}
where ${a_1}^2 + {a_2}^2 + {b_1}^2 + {b_2}^2 = 1$, ${\rm{\alpha }} \in {\mathbb{R}}$.\\

\indent According to Eq. (\ref{Eq:U}), it is easy to calculate and obtain Eq. (\ref{Eq:Cii(U)}).
\begin{equation}
\left\{ {\begin{array}{*{20}{c}}
  {{C_{11}}\left( {\mathbf{U}} \right) = {\text{ }}2\left( {c - d} \right) - 1} \\
  {{C_{22}}\left( {\mathbf{U}} \right) = {\text{ }}2\left( {c + d} \right) - 1} \\
  {{C_{33}}\left( {\mathbf{U}} \right) = {\text{ }}2\left( {a_1^2 + a_2^2} \right) - 1}
\end{array}} \right.\label{Eq:Cii(U)}
\end{equation}
\indent where $c = \frac{1}{2} + {a_1}  {a_2}  sin{\kern 1pt} {\kern 1pt} \alpha  + \frac{{cos{\kern 1pt} {\kern 1pt} \alpha }}{2}({a_1}^2 - {a_2}^2)$,\\
$d = {b_1}  {b_2}  sin{\kern 1pt} {\kern 1pt} \alpha  + \frac{{cos{\kern 1pt} {\kern 1pt} \alpha }}{2}({b_1}^2 - {b_2}^2)$.\\
\indent If the solutions ${p_{{j}}}\left( {j = 0, \ldots ,3} \right)$ of Eq. (\ref{4.1}) exist, and ${p_{{j}}} \ge 0\left( {j = 0, \ldots ,3} \right)$, then lemma 5 is  proved.
\begin{equation}
\left\{ {\begin{array}{*{20}{c}}
{{{\bf{P}}}\left( {\bf{U}} \right) = \mathop \sum \limits_{{{j}} = 0}^3 {p_{{j}}}  {{\bf{P}}}\left( {{\sigma _{{j}}}} \right)}\\
{\mathop \sum \limits_{{{j}} = 0}^3 {p_{{j}}} = 1}
\end{array}} \right.\label{4.1}
\end{equation}
\indent The only solution of Eq. (\ref{4.1}) is as follows.

\begin{equation}
\left\{ {\begin{array}{*{20}{c}}
  {{p_0} = \frac{1}{4}[{C_{11}}\left( {\mathbf{U}} \right) + {C_{22}}\left( {\mathbf{U}} \right) + {C_{33}}\left( {\mathbf{U}} \right) + 1]} \\
  {{p_1} = \frac{1}{4}[{C_{11}}\left( {\mathbf{U}} \right) - {C_{22}}\left( {\mathbf{U}} \right) - {C_{33}}\left( {\mathbf{U}} \right) + 1]} \\
  {{p_2} = \frac{1}{4}[ - {C_{11}}\left( {\mathbf{U}} \right) + {C_{22}}\left( {\mathbf{U}} \right) - {C_{33}}\left( {\mathbf{U}} \right) + 1]} \\
  {{p_3} = \frac{1}{4}[ - {C_{11}}\left( {\mathbf{U}} \right) - {C_{22}}\left( {\mathbf{U}} \right) + {C_{33}}\left( {\mathbf{U}} \right) + 1]}
\end{array}} \right.
\end{equation}

\indent Furthermore, according to Eq. (\ref{Eq:Cii(U)}), it is easy to prove ${p_{{j}}} \ge 0\left( {j = 0, \ldots ,3} \right)$.
\begin{equation}
\left\{ {\begin{array}{*{20}{c}}
{{p_0}^ = \frac{1}{2}{{\left[ {{a_1}\sqrt {\left( {1 + \cos \alpha } \right)}  \pm {a_2}\sqrt {\left( {1 - \cos {\kern 1pt} {\kern 1pt} \alpha } \right)} } \right]}^2} \ge 0}\\
{{p_1}^ = \frac{1}{2}{{\left[ {{b_1}\sqrt {\left( {1 - \cos \alpha } \right)}  \pm {b_2}\sqrt {\left( {1 + \cos {\kern 1pt} {\kern 1pt} \alpha } \right)} } \right]}^2} \ge 0}\\
{{p_2}^ = \frac{1}{2}{{\left[ {{b_1}\sqrt {\left( {1 + \cos \alpha } \right)}  \pm {b_2}\sqrt {\left( {1 - \cos {\kern 1pt} {\kern 1pt} \alpha } \right)} } \right]}^2} \ge 0}\\
{{p_3}^ = \frac{1}{2}{{\left[ {{a_1}\sqrt {\left( {1 - \cos \alpha } \right)}  \pm {a_2}\sqrt {\left( {1 + \cos {\kern 1pt} {\kern 1pt} \alpha } \right)} } \right]}^2} \ge 0}
\end{array}} \right.
\end{equation}
\end{proof}
\indent Lemma 5 just illustrates that for an arbitrary ${\bf{U}}$, ${{\bf{P}}}\left( {\bf{U}} \right)$ corresponds to a point in the regular tetrahedron ${\bf{T}}_{\rm{DC}}$ with four vertices ${{\bf{P}}}\left( {{\sigma _j}} \right)\left( {j = 0, \ldots ,3} \right)$, as shown in Fig. \ref{figure_2}. Next, lemma 6 (i.e. the inverse proposition of lemma 5) will illustrate that for a point in ${\bf{T}}_{\rm{DC}}$, there must exist ${\bf{U}} \in {\bf{U}}\left( 2 \right)$ such that ${{\bf{P}}}\left( {\bf{U}} \right)$ corresponds to this point. Now, lemma 6 is proved as follows.\\

\indent {\bf{\emph{Lemma 6}.}} $\forall {p_{{j}}} \ge 0,\mathop \sum \limits_{{{j}} = 0}^3 {p_{{j}}} = 1$, there must exist ${\bf{U}} \in {\bf{U}}\left( 2 \right)$ such that ${{\bf{P}}}\left( {\bf{U}} \right) = \mathop \sum \limits_{{{j}} = 0}^3 {p_{{j}}}  {{\bf{P}}}\left( {{\sigma _{{j}}}} \right)$.
\begin{proof}
\indent Eq. (\ref{Eq:U}) shows that ${\bf{U}}$ is a unitary matrix regardless of
$\alpha $, here, let $\alpha  = 2k\pi $, where $k \in \mathbb{Z} $. And Eq. (\ref{Eq:Cii(U)}) shows the detailed representation of ${\bf{P}}( {\bf{U}})$. Given a set of ${p_{{j}}}$ $({p_{{j}}}\ge 0,\mathop \sum \limits_{{{j}} = 0}^3 {p_{{j}}} = 1,j=0,\ldots,3$), only if solutions ${a_1},{a_2},{b_1},{b_2}$ of Eq. (\ref{Eqs:4.2}) exist, is the lemma 6 proved.
\begin{equation}
\left\{ {\begin{array}{*{20}{c}}
{{{\bf{P}}}\left( {\bf{U}} \right) = \mathop \sum \limits_{{{j}} = 0}^3 {p_{{j}}}  {{\bf{P}}}\left( {{\sigma _{{j}}}} \right)}\\
{{a_1}^2 + {a_2}^2 + {b_1}^2 + {b_2}^2 = 1}
\end{array}} \right.\label{Eqs:4.2}
\end{equation}

Solutions of Eq. (\ref{Eqs:4.2}) are easy to be obtained as follows.

\begin{equation}
\left\{ {\begin{array}{*{20}{c}}
{{a_1} =  \pm \sqrt {{p_0}} }\\
{{a_2} =  \pm \sqrt {{p_3}} }\\
{{b_1} =  \pm \sqrt {{p_2}} }\\
{{b_2} =  \pm \sqrt {{p_1}} }
\end{array}} \right.
\end{equation}
\end{proof}

\indent We aim to find the infimum  $\inf {\kern 1pt} {C_{\rm{DC}}}$. Lemmas 5 and 6 illustrate that for all ${\bf{U}} \in {\bf{U}}(2)$, ${\bf{P}}({\bf{U}})$ forms the regular tetrahedron ${{\bf{T}}_{\rm{DC}}}$. Therefore, $\mathop {\inf }\limits_{{\kern 1pt} {\bf{U}} \in {\bf{U}}(2)} {\kern 1pt} {\kern 1pt} {C_{\rm{DC}}} = \mathop {\min }\limits_{ {{{\bf{T}}_{\rm{DC}}}}} {\kern 1pt} {C_{\rm{DC}}}$. Now $\mathop {\min }\limits_{ {{{\bf{T}}_{\rm{DC}}}}} {C_{\rm{DC}}}$ is calculated as follows.

\indent {\bf{\emph{Theorem 2}.}} $\forall {\kern 1pt} {\bf{U}} \in {\bf{U}}\left( 2 \right),{C_{\rm{DC}}}\left( {\bf{U}} \right) \in \left[ { - \frac{1}{{27}},1} \right]$.\\
\begin{proof}
According to lemma 5, ${\bf{P}}({\sigma _{{j}}})=$ $({C_{11}}( {{\sigma _{{j}}}}),{C_{22}}( {{\sigma _{{j}}}}),{C_{33}}( {{\sigma _{{j}}}}))'$ $( {j = 0, \ldots ,3})$ are four vertices of ${{\bf{T}}_{\rm{DC}}}$. Therefore, $\mathop {\min }\limits_{ {{{\bf{T}}_{\rm{DC}}}}} {C_{\rm{DC}}}$ is equivalent to Eq. (\ref{Eq:kkt2}).
\begin{equation}
\begin{array}{l}
\mathop {\min }\limits_{ {{{\bf{T}}_{\rm{DC}}}}} {C_{\rm{DC}}} = \prod\limits_{i = 1}^3 {\sum\limits_{j = 0}^3 {{p_{{j}}}  {C_{ii}}\left( {{\sigma _{{j}}}} \right)} } \\
{\kern 1pt} {\kern 1pt} {\kern 1pt} {\kern 1pt} {\kern 1pt} s.t.{\kern 1pt} {\kern 1pt} {\kern 1pt} {\kern 1pt} {\kern 1pt} {\kern 1pt} {\kern 1pt} {\kern 1pt} {\kern 1pt} {\kern 1pt} {\kern 1pt} {\kern 1pt} {\kern 1pt} {\kern 1pt} {\kern 1pt} {\kern 1pt} {\kern 1pt} {\kern 1pt} {\kern 1pt} {\kern 1pt} {\kern 1pt} {\kern 1pt} {\kern 1pt} {\kern 1pt} {\kern 1pt} {\kern 1pt} {\kern 1pt} \sum\limits_{j = 0}^3 {{p_{{j}}} = 1}
\end{array}\label{Eq:kkt2}
\end{equation}
\indent A Lagrangian function is constructed as follows.
\begin{equation}
F({p_j},\lambda ) = \prod\limits_{i = 1}^3 {\sum\limits_{j = 0}^3 { {{p_{{j}}}  {C_{ii}}\left( {{\sigma _{{j}}}} \right)} } }  - \lambda \left( {\mathop \sum \limits_{{{j}} = 0}^3 {p_{{j}}} - 1} \right) .\label{Eq:F(p,lambda)}
\end{equation}
\indent A total of 92 extreme points (see Supplemental Material \cite{Authors12}) are obtained with KKT conditions, and $\mathop {\min }\limits_{ {{{\bf{T}}_{\rm{DC}}}}} {C_{\rm{DC}}} =  - \frac{1}{{27}}$. Hence, ${C_{\rm{DC}}}\left( {\bf{U}} \right) \ge  - \frac{1}{{27}}$.
\end{proof}

\section{A MIXTURE OF QUANTUM COMMON CAUSES AND QUANTUM CAUSALITY }\label{sec:IV}
\indent The mixture of quantum common causes and quantum causality means that two qubits come from the $p$-mixture of  quantum common causes ($\rho$) and quantum causality ($\bf{U}$), where $p$ is the probability of quantum common causes, which corresponds to the right one in Fig. \ref{figure_1}. In order to discriminate quantum common causes, quantum causality and a combination of both in general, first, the vector-valued function ${{\bf{P}}}\left( {{\bf{\rho ,U},\emph{p}}} \right)$ is defined as follows.\\
\indent \emph{Definition 3.} The vector-valued function ${{{\bf{P}}}\left( {\rho ,{\bf{U}}},\emph{p} \right)}$ on the $p$-mixture of density operator $\rho $ and the unitary matrix ${\bf{U}}$ is defined as
\begin{equation}
{{\bf{P}}}\left( {\rho ,{\bf{U}},p}\right) \equiv \left( {\begin{array}{*{20}{c}}
{{C_{11}}\left( {\rho ,{\bf{U}}},p \right)}\\
{{C_{22}}\left( {\rho ,{\bf{U}}},p \right)}\\
{{C_{33}}\left( {\rho ,{\bf{U}}},p \right)}
\end{array}} \right) .
\end{equation}
\indent The following theorem shows that ${{\bf{P}}}\left( {\rho ,{\bf{U}},p} \right)$ corresponds to a probabilistic mixture of two points in ${\bf{T}}_{\rm{CC}}$ and ${\bf{T}}_{\rm{DC}}$, respectively. \\
\begin{figure}[!htp]
 \centering
 \includegraphics[width=2.4in,height=1.5in]{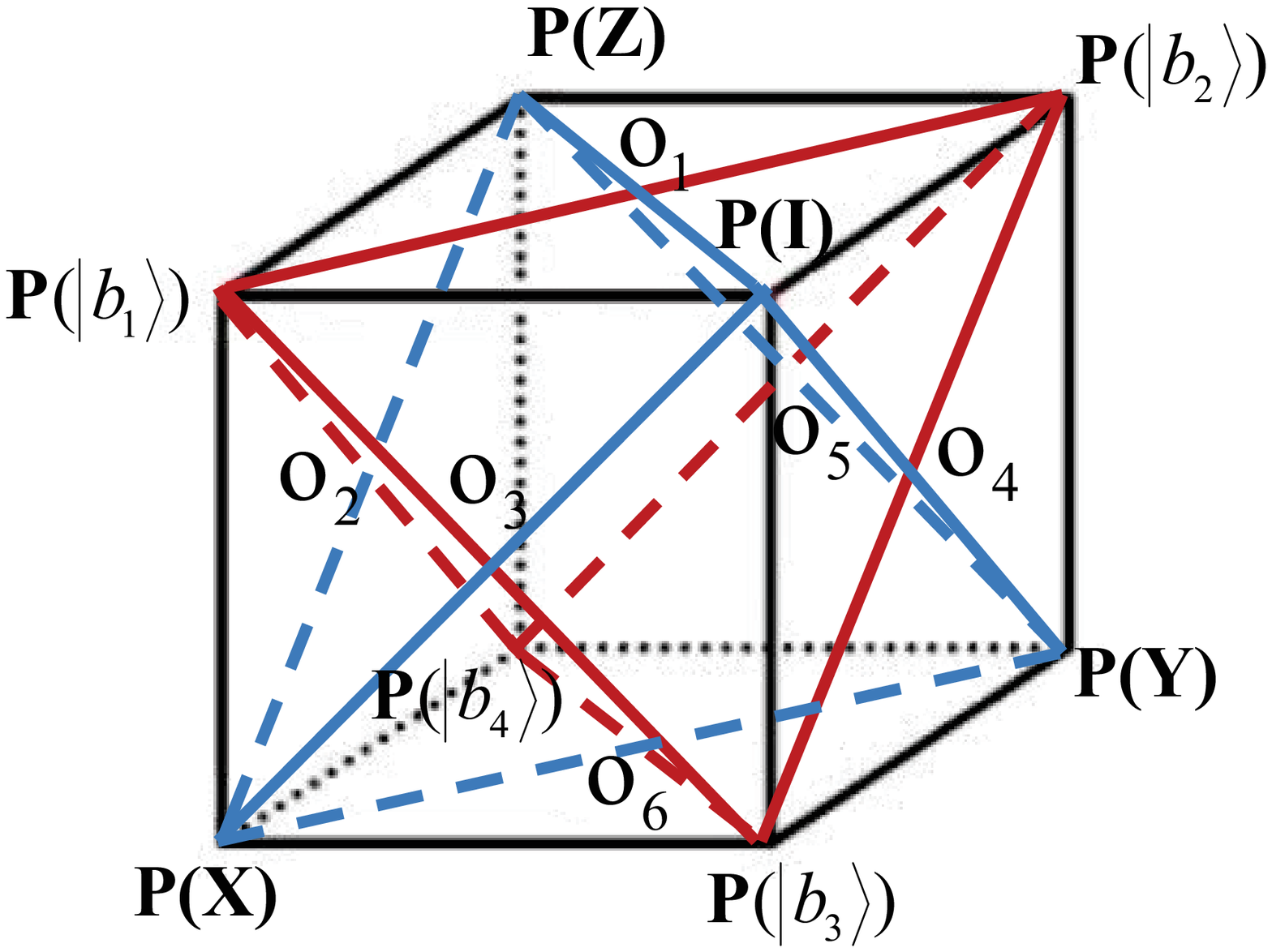}
\caption{\label{figure_2} The geometric interpretation of quantum common causes, quantum causality and a combination of both. Take the center of the cube as the origin, the x, y and z axes are parallel to the sides ${\bf{P}}(|{b_4}\rangle ) {\bf{P}}(\bf{X}) $, ${\bf{P}}(|{b_4}\rangle ){\bf{P}}(\bf{Y}) $, $ {\bf{P}}(|{b_4}\rangle){\bf{P}}(\bf{Z} )$, and the positive directions point to ${\bf{P}}(\bf{X})$, ${\bf{P}}(\bf{Y})$, ${\bf{P}}(\bf{Z} )$ respectively. The red regular tetrahedron ${\bf{T}}_{\rm{CC}}$ consists of ${{\bf{P}}}\left( {|\varphi \rangle } \right)$. The blue regular tetrahedron ${\bf{T}}_{\rm{DC}}$ consists of ${{\bf{P}}}\left( {\bf{U}} \right)$.  ${{\rm{O}}_{i}}(i=1, \ldots, 6)$ are the central points of the six faces of the cube. The overlapped area $\bf{O}$ of ${\bf{T}}_{\rm{CC}}$ and ${\bf{T}}_{\rm{DC}}$ is shown in Fig. \ref{figure_3}.}
\end{figure}

\indent {\bf{\emph{Theorem 3}.}} For $\forall \rho \in {\mathbb{C}}^{4 \times 4}, \forall {\bf{U}} \in {\bf{U}}(2)$, $p \in \left[ {0,1} \right]$, ${{{\bf{P}}}\left( {\rho ,{\bf{U}},p} \right)} = p{{\bf{P}}}\left( \rho  \right) + \left( {1 - p} \right){{\bf{P}}}\left( {\bf{U}} \right)$ holds, where ${{\bf{P}}}\left( \rho  \right) \in {{{\bf{T}}_{\rm{CC}}}}$, ${{\bf{P}}}\left( {\bf{U}} \right) \in {{{\bf{T}}_{\rm{DC}}}}$.

\begin{proof}
\indent If $p = 1 $ or $p = 0$, then ${{{\bf{P}}}\left( {\rho ,{\bf{U}},p} \right)} \in {{\bf{T}}_{\rm{CC}}}$ or ${{{\bf{P}}}\left( {\rho ,{\bf{U}},p} \right)}\in {{\bf{T}}_{\rm{DC}}} $. Theorem 3 holds immediately. Therefore, it is only necessary to prove theorem 3 holds when $p \in (0,1)$.\\
\indent When $p \in (0,1)$, according to Eq.(\ref{Eq:C(rho)}), apparently, any point ${{{\bf{P}}}\left( {\rho ,{\bf{U}},p} \right)} \in \bf{D}$, where ${\bf{D}}$ is a regular hexahedron with vertices ${{\bf{P}}}(|{b_j}\rangle )\left( {j = 1, \ldots ,4} \right)$, ${{\bf{P}}}\left( {{\sigma _k}} \right)\left( {k = 0, \ldots ,3} \right)$, see Fig. \ref{figure_2}. Since that $\bf{D}$ is a convex set. Hence,
\begin{equation}
{{{\bf{P}}}\left( {\rho ,{\bf{U}},p} \right)} = \mathop \sum \limits_{{{j}} = 1}^4 \left[ {{p_{{j}}} {{\bf{P}}}\left( {|{b _{{j}}}\rangle} \right)} \right] + \sum\limits_{k = 5}^8 {\left[ {{p_k}{{\bf{P}}}\left( {{\sigma_{k - 5}} } \right)} \right]}\label{Cii(rho,u)},
\end{equation}
where ${p_{{j}}} \ge 0\left( {j = 1, \ldots ,4} \right)$, ${p_k} \ge 0\left( {k = 5, \ldots ,8} \right)$, and $\mathop \sum \limits_{{{j}} = 1}^4 {p_{{j}}} + \mathop \sum \limits_{k = 5}^8 {p_k} = 1$.\\
\indent Let $p = \sum\limits_{j = 1}^4 {{p_{{j}}}}$, ${q_j} = \frac{{{p_{{j}}}}}{{p}} \ge 0 \left( {j = 1, \ldots ,4} \right)$. ${q_k} = \frac{{{p_k}}}{{1-p}} \ge 0 \left( {k = 5, \ldots ,8} \right)$. Then $\mathop \sum \limits_{{{j}} = 1}^4 {q_{{j}}} = 1$, $\mathop \sum \limits_{k = 5}^8 {q_k} = 1$. \\
\indent Eq. (\ref{Cii(rho,u)}) is equivalent to Eq. (\ref{Eq:C(rho,u)}).
\begin{equation}
\begin{split}
{{{\bf{P}}}\left( {\rho ,{\bf{U}},p} \right)} &= p\mathop \sum \limits_{{{j}} = 1}^4 \left[ {{q_j}  {{\bf{P}}}\left( {|{ b_{{j}}}\rangle} \right)} \right] + \left( {1 - p} \right)\sum\limits_{k = 5}^8 {\left[ {{q_k}  {{\bf{P}}}\left( {{\sigma_{k - 5}} } \right)} \right]}  \\
&= p{{\bf{P}}}\left( \rho \right) + \left( {1 - p} \right){{\bf{P}}}\left( {\bf{U}}  \right),
\end{split}\label{Eq:C(rho,u)}
\end{equation}
where ${{\bf{P}}}\left( \rho  \right) \in {{{\bf{T}}_{\rm{CC}}}}$, ${{\bf{P}}}\left( {\bf{U}} \right) \in {{{\bf{T}}_{\rm{DC}}}}$, $p \in \left( {0,1} \right)$.

\end{proof}

\section{Quantum Common Causes and Quantum Causality in the Overlapped Area}\label{sec:V}
\begin{figure}[!htp]
 \centering
 \includegraphics[width=2.2in,height=1.8in]{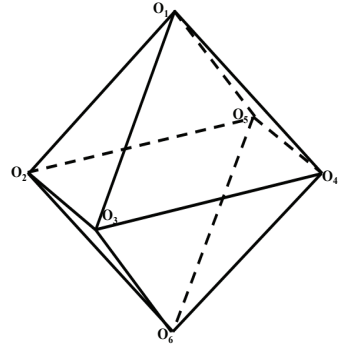}
\caption{\label{figure_3} Overlapped area $\bf{O}$ of ${\bf{T}}_{\rm{CC}}$ and ${\bf{T}}_{\rm{DC}}$. The corresponding six vertices are ${{\rm{O}}_{i}}(i=1, \ldots, 6)$, respectively, which are the central points of the six faces of the cube in Fig. \ref{figure_2}.}
\end{figure}

\indent Fig. \ref{figure_2} implies that quantum common causes and quantum causality can not be discriminated in the overlapped area $\bf{O}$ of ${\bf{T}}_{\rm{CC}}$ and ${\bf{T}}_{\rm{DC}}$ by ${\bf{P}}$ ($\bf{O}$ is shown in Fig. \ref{figure_3}). To discriminate quantum common causes and quantum causality more completely, a heuristic principle is presented in this section: we try to find a new vector-valued function ${\bf{P}'}$ such that quantum common causes and quantum causality in $\bf{O}$ can be distinguished, at least to some extent. Then the combination of ${\bf{P}}$ and ${\bf{P}'}$ can more effectively identify quantum common causes and quantum causality. In general, ${\bf{P}'}$ can be constructed via transforming the basis vectors of project measurements, i.e., $|{x_0}\rangle$, $|{x_1}\rangle$, $|{y_0}\rangle$, $|{y_1}\rangle$, $|{z_0}\rangle$, and $|{z_1}\rangle$, by an appropriate unitary transformation $\bf{V}$. In the following part, some theoretical observations and simulation results are given to facilitate the above heuristic principle. Specifically,Theorems 4 and 5 show how to connect $\bf{P}'$ under the transformed basis vectors to $\bf{P}$ under the original basis vectors. \\

\indent {\bf{\emph{Theorem 4}.}} For $\forall \rho \in {\mathbb{C}}^{4 \times 4}$, if the ${\bf{P}}'(\rho)$ is constructed by the effect of a unitary transformation $\bf{V}$ on the basis vectors of project measurements with measurement operators ${\sigma _i}\otimes {\sigma _i} (i = 1, \ldots ,3)$, then ${\bf{P}}'(\rho) = {\bf{P}}((\bf{V} \otimes \bf{V})^{'} \rho(\bf{V} \otimes \bf{V}))$, where ${\bf{V}} \in {\bf{U}}(2)$.

\begin{proof}
Let ${{\bf{P}}'}(\rho ) \equiv \left( {\begin{array}{*{20}{c}}
{{C_{11}}'(\rho )}\\
{{C_{22}}'(\rho )}\\
{{C_{33}}'(\rho )}
\end{array}} \right)$. According to Eq.(1), ${C_{ii}}'(\rho ) \equiv {p'}(k = m|ii) - {p'}(k \ne m|ii)$$\left( {i = 1, \ldots ,3} \right)$, where ${p'}(k = m|ii)$ and ${p'}(k \ne m|ii)$ are the probabilities of the same measurement result and the different measurement results that the transformed observable $({\bf{V}}{\sigma _i}{\bf{V}}^{'}) \otimes ({\bf{V}}{\sigma _i}{\bf{V}}^{'})$ is used on the $\rho$, where ${\bf{V}} \in {\bf{U}}(2)$. When $\rho$ is a pure state, it is equivalent to define ${C_{ii}}'(\left| \varphi  \right\rangle )$ on the state vector $\left| \varphi  \right\rangle$, where $\left| \varphi  \right\rangle \in {\mathbb{C}^4}$. That is to prove ${C_{ii}}^{'}(\left| \varphi  \right\rangle ) = {C_{ii}}({({\bf{V}} \otimes {\bf{V}})^{'}}  \left| \varphi  \right\rangle )$$\left( {i = 1, \ldots ,3} \right)$.\\
\begin{widetext}
\begin{eqnarray}
\begin{array}{l}
{C_{ii}}^{'}(\left| \varphi  \right\rangle ) \equiv {p^{'}}(k = m|ii) - {p^{'}}(k \ne m|ii)\\
=2 p^{'}(k = m|ii) - 1\\
 = 2\{ \left\langle \varphi  \right|({\bf{V}} \left| {{m_0}} \right\rangle ) \otimes ({\bf{V}} \left| {{m_0}} \right\rangle ) {[({\bf{V}}  \left| {{m_0}} \right\rangle ) \otimes ({\bf{V}}  \left| {{m_0}} \right\rangle )]^{'}}\left| \varphi  \right\rangle {\kern 1pt}  + \left\langle \varphi  \right|({\bf{V}}  \left| {{m_{\rm{1}}}} \right\rangle ) \otimes ({\bf{V}}  \left| {{m_{\rm{1}}}} \right\rangle )  {[({\bf{V}} \left| {{m_{\rm{1}}}} \right\rangle ) \otimes ({\bf{V}}  \left| {{m_{\rm{1}}}} \right\rangle )]^{'}}\left| \varphi  \right\rangle \}  - 1\\
 = 2\{ \left\langle \varphi  \right|  ({\bf{V}} \otimes {\bf{V}})  \left| {{m_{\rm{0}}}{m_0}} \right\rangle \left\langle {{m_0}{m_0}} \right|  {({\bf{V}} \otimes {\bf{V}})^{'}}  \left| \varphi  \right\rangle  + \left\langle \varphi  \right|  ({\bf{V}} \otimes {\bf{V}}) \left| {{m_{\rm{1}}}{m_{\rm{1}}}} \right\rangle \left\langle {{m_1}{m_1}} \right|  {({\bf{V}} \otimes {\bf{V}})^{'}} \left| \varphi  \right\rangle \}  - 1\\
 = {C_{ii}}({({\bf{V}} \otimes {\bf{V}})^{'}}  \left| \varphi  \right\rangle )
\end{array}\label{long_eq1}
\end{eqnarray}
\end{widetext}
where $|\left. {{m_0}} \right\rangle $ and $|\left. {{m_1}} \right\rangle $ represent two eigenstates of ${\sigma _i}(i = 1, \ldots ,3)$.\\
\indent When $\rho$ is a mixed state, $\rho {\rm{ = }}\sum\limits_{j = 1}^n {{p_j}\left| {{\varphi _j}} \right\rangle \left\langle {{\varphi _j}} \right|}  $, where $\sum\limits_{j = 1}^n {{p_j}}=1  $, it can be regarded as a convex combination of $n$ pure states. For each pure state $\left| {{\varphi _j}} \right\rangle$, Eq.(\ref{long_eq1}) holds. Therefore, ${C_{ii}}'(\rho ) = {C_{ii}}(({\bf{V}} \otimes {\bf{V}})'\rho({\bf{V}} \otimes {\bf{V}}))\left( {i = 1, \ldots ,3} \right)$ holds.
\end{proof}

According to theorem 4 and lemma 1, clearly, for all $\rho \in {\mathbb{C}}^{4 \times 4}$, $\bf{P}'(\rho)$ forms the regular tetrahedron ${{\bf{T}}_{\rm{CC}}}$. Therefore, for a $\rho$ in the overlapped area $\bf{O}$, if there exists a unitary matrix $\bf{V} \in {{\bf{U}}}(2)$ such that ${\bf{P}}'(\rho) = {\bf{P}}((\bf{V} \otimes \bf{V})^{'} \rho({\bf{V}} \otimes {\bf{V}})) \notin \bf{O}$, then the $\rho$ is rotated to ${\bf{T}}_{\rm{CC}}{\bf{/}}{\kern 1pt} {\bf{O}}$ via the unitary matrix $\bf{V}$. That is to say, this case is discriminated.\\
\begin{table*}[!htp]
 \scalebox{1.1}{
 \begin{tabular}{|c|c|c|}
 \hline
 Unitary matrix&Proportion of&Proportion of\\
 $\bf{V}$&${\bf{P}}(\rho ) \in {\bf{O}} \to {{\bf{P}}'}(\rho ) \in {\bf{OTC/O}}$&${\bf{P}}(\bf{U} ) \in {\bf{O}} \to {{\bf{P}}'}(\bf{U} ) \in {\bf{OTD/O}}$\\

 \hline
 ${\bf{V}}_{1}  \approx \left( {\begin{array}{*{20}{c}}
{\begin{array}{*{20}{c}}
{0.1813{\rm{ }} - {\rm{ }}0.5744i}\\
{ - 0.6807{\rm{ }} + {\rm{ }}0.4170i}
\end{array}}&{\begin{array}{*{20}{c}}
{0.2656{\rm{ }} + {\rm{ }}0.7527i}\\
{ - 0.2213{\rm{ }} + {\rm{ }}0.5602i}
\end{array}}
\end{array}} \right)$&36.44\%&58.91\%\\
 \hline
 $\bf{V}_{2} \approx\left( {\begin{array}{*{20}{c}}
{\begin{array}{*{20}{c}}
{- 0.1080{\rm{ }} + {\rm{ }}0.7959i}\\
{- 0.4763{\rm{ }} - {\rm{ }}0.3577i}
\end{array}}&{\begin{array}{*{20}{c}}
{ 0.4848{\rm{ }} - {\rm{ }}0.3461i}\\
{ - 0.0888{\rm{ }} - {\rm{ }}0.7983i}
\end{array}}
\end{array}} \right)$&35.84\%&57.32\%\\
 \hline
 ${{\bf{V}}_3}  \approx \left( {\begin{array}{*{20}{c}}
{\begin{array}{*{20}{c}}
{{\rm{ - 0}}{\rm{.2947  +  0}}{\rm{.5266}}i}\\
{{\rm{ - 0}}{\rm{.6926  -  0}}{\rm{.3950}}i}
\end{array}}&{\begin{array}{*{20}{c}}
{{\rm{0}}{\rm{.7483  -  0}}{\rm{.2754}}i}\\
{{\rm{ - 0}}{\rm{.2039  -  0}}{\rm{.5680}}i}
\end{array}}
\end{array}} \right)$&29.9\%&50.64\%\\
 \hline
 ${\bf{V}}_{4}  \approx \left( \begin{array}{l}
{\rm{0}}{\rm{.3482  +  0}}{\rm{.3352}}i \quad \quad  {\rm{      - 0}}{\rm{.3442  +  0}}{\rm{.8050}}i\\
{\rm{ - 0}}{\rm{.2069  +  0}}{\rm{.8507}}i \quad \quad {\rm{     0}}{\rm{.4796  -  0}}{\rm{.0597}}i
\end{array} \right)$&33.45\% &52.56\%\\
 \hline

 \end{tabular}}
 \caption{\label{table:2} Proportion of the overlapped area $\bf{O}$ transferred to the distinguishable area $\bf{OTC/O}(\bf{OTD/O})$ via the specific $\bf{V}$. Randomly generate 20000 $\rho$ ($\bf{U}$) meeting $\bf{P}(\rho) \in \bf{O}$ ($\bf{P}(U) \in \bf{O}$), and calculate the proportion of $\rho(\bf{U})$ transformed to $\bf{OTC/O}(\bf{OTD}{\bf{/}}{\kern 1pt} {\bf{O}})$. For ${\bf{V}_{1}}$ and ${\bf{V}_{2}}$, they could make $\bf{O}$ be converted to $\bf{OTC}$ via the single ${\bf{V}_{1}}$ or ${\bf{V}_{2}}$. However, $\bf{O}$ could not be converted to $\bf{OTC}$ via the single ${\bf{V}_{3}}$ or ${\bf{V}_{4}}$. This might be the reason why transfer ratios of the second column under the effect of ${\bf{V}_{1}}$ and ${\bf{V}_{2}}$ are larger than the counterparts of ${\bf{V}_{3}}$ and ${\bf{V}_{4}}$. $\bf{O}$ could be converted to $\bf{OTD}$ via the single ${\bf{V}_{1}}$, ${\bf{V}_{2}}$, ${\bf{V}_{4}}$, but $\bf{O}$ could not be converted to $\bf{OTD}$ via the single ${\bf{V}_{3}}$. This might be the reason why transfer ratio of the third column under the effect of ${\bf{V}_{3}}$ is lower than the counterparts of ${\bf{V}_{1}}$, ${\bf{V}_{2}}$ and ${\bf{V}_{4}}$.}
\end{table*} 
Based on the above theoretical observation, some simulation experiments of the overlapped area $\bf{O}$ were carried out. First, the simulation results show that $\bf{O}$ can be converted to $\bf{OTC}$ under the effect of all appropriate unitary transformations $\bf{V}$, where $\bf{OTC}$ is the regular tetrahedron ${{\bf{T}}_{\rm{CC}}}$ from which is dug out a small tetrahedron with vertices $\bf{P}(|{b_3}\rangle),{{\rm{O}}_4},{{\rm{O}}_5},{{\rm{O}}_6}$. The reason why $\bf{OTC}$ does not include the small tetrahedron is that the form of $(\bf{V} \otimes \bf{V})^{'}\rho ({\bf{V}} \otimes {\bf{V}})$ limits the resulted $4\times4$ density matrix. Interestingly, there exists some special unitary matrices $\bf{V}$, for example, ${\bf{V}}  \approx \left( {\begin{array}{*{20}{c}}
{\begin{array}{*{20}{c}}
{0.1813{\rm{ }} - {\rm{ }}0.5744i}\\
{ - 0.6807{\rm{ }} + {\rm{ }}0.4170i}
\end{array}}&{\begin{array}{*{20}{c}}
{0.2656{\rm{ }} + {\rm{ }}0.7527i}\\
{ - 0.2213{\rm{ }} + {\rm{ }}0.5602i}
\end{array}}
\end{array}} \right)$, such that the overlapped area $\bf{O}$ is converted to $\bf{OTC}$ by the single unitary matrix $\bf{V}$. Although it does not mean that any unitary matrix can make $\bf{O}$ be transferred to $\bf{OTC}$. For more specific instructions, the proportion that $\rho$ in $\bf{O}$ is transformed to $\bf{OTC}{\bf{/}}{\kern 1pt} {\bf{O}}$ is investigated via some specific $\bf{V}$, the simulation results are shown in Table \ref{table:2}. \\

\indent {\bf{\emph{Theorem 5}.}} For $\forall \bf{U} \in {\bf{U}}(2)$, if ${\bf{P}}'(\bf{U})$ is constructed by the effect of a unitary transformation $\bf{V}$ on the basis vectors of project measurements with measurement operators ${\sigma _i}(i = 1, \ldots ,3)$, then ${{\bf{P}}^{'}}({\bf{U}}) = {\bf{P}}({{\bf{V}}^{'}}{\bf{UV}})$, where ${\bf{V}} \in {\bf{U}}(2)$.

\begin{proof}
Let ${{\bf{P}}'}({\bf{U}}) \equiv \left( {\begin{array}{*{20}{c}}
{{C_{11}}'({\bf{U}})}\\
{{C_{22}}'({\bf{U}})}\\
{{C_{33}}'({\bf{U}})}
\end{array}} \right)$. According to Eq.(\ref{def:C(U)}), it is only necessary to prove that ${C_{ii}}'({\bf{U}}) = {C_{ii}}({{\bf{V}}'}  {\bf{U}}  {\bf{V}})(i = 1, \ldots ,3)$. According to Eq.(\ref{C_11}), Eq.(\ref{Eq:Cii'_Cii}) is derived as follows.
\begin{equation}
\begin{array}{l}
{C_{ii}}'({\bf{U}}) = 2{p_{B|A}}'(\left| {{m_0}} \right\rangle ) - 1\\
 = 2{p_{B|A}}({\bf{V}}\left| {{m_0}} \right\rangle ) - 1\\
 = 2{({\bf{U}} {\bf{V}}  \left| {{m_0}} \right\rangle )'}  {\bf{V}}  \left| {{m_0}} \right\rangle \left\langle {{m_0}} \right|  {{\bf{V}}'}  {\bf{U}}  {\bf{V}} \left| {{m_0}} \right\rangle  - 1\\
 = \left\langle {{m_0}} \right|  {{\bf{V}}'}  {{\bf{U}}'}  {\bf{V}}  \left| {{m_0}} \right\rangle \left\langle {{m_0}} \right|  {{\bf{V}}'} {\bf{U}}  {\bf{V}}  \left| {{m_0}} \right\rangle  - 1\\
 = {C_{ii}}({{\bf{V}}'}  {\bf{U}}  {\bf{V}})
\end{array},\label{Eq:Cii'_Cii}
\end{equation}
where ${p_{B|A}}({\bf{V}}\left| {m_0} \right\rangle )$ represents the probability that $B$ is collapsed to ${\bf{V}} \left| {{m_0}} \right\rangle$ under the condition that $A$ is collapsed to ${\bf{V}} \left| {{m_0}} \right\rangle$ measured by ${\bf{V}} \left| {{m_0}} \right\rangle$. $\left| {{m_0}} \right\rangle $ represents one of two eigenstates of Pauli matrices ${\sigma _i}(i = 1, \ldots ,3)$.
\end{proof}

According to theorem 5 and lemma 5, obviously, for all $\bf{U} \in {\bf{U}}(2)$, $\bf{P}'(\bf{U})$ forms the regular tetrahedron ${{\bf{T}}_{\rm{DC}}}$. Hence, for a $\bf{U}$ in $\bf{O}$, if there exists $\bf{V} \in {\bf{U}}(2)$ such that ${{\bf{P}}^{'}}({\bf{U}}) = {\bf{P}}({{\bf{V}}^{'}}{\bf{UV}}) \notin \bf{O}$, then $\bf{U}$ is shifted out to ${\bf{T}}_{\rm{DC}}{\bf{/}}{\kern 1pt} {\bf{O}}$ via the unitary matrix $\bf{V}$. This case is discriminated.\\
\indent In terms of simulations, simulation results illustrate that $\bf{O}$ is transformed to $\bf{OTD}$ under the effect of all appropriate unitary transformations $\bf{V}$, where $\bf{OTD}$ is the regular tetrahedron ${{\bf{T}}_{\rm{DC}}}$ from which is dug out a small tetrahedron with vertices ${{\rm{O}}_1},{{\rm{O}}_2},{{\rm{O}}_3},\bf{P}(\bf{Z})$. The reason why $\bf{OTD}$ is dug out the small tetrahedron is that $\bf{{V}'UV}$ limits the arbitrariness of the resulted $2 \times 2$ unitary matrix. Interestingly, there exists some special unitary matrices $\bf{V}$, for example, ${\bf{V}}  \approx \left( {\begin{array}{*{20}{l}}
{{\rm{0}}.{\rm{3482 + 0}}.{\rm{3352}}i\quad \quad {\rm{ - 0}}.{\rm{3442 + 0}}.{\rm{8050}}i}\\
{{\rm{ - 0}}.{\rm{2069 + 0}}.{\rm{8507}}i\quad \quad {\rm{0}}.{\rm{4796 - 0}}.{\rm{0597}}i}
\end{array}} \right)$, such that the overlapped area $\bf{O}$ is converted to $\bf{OTD}$ by the single unitary matrix $\bf{V}$. However, it does not mean that any unitary matrix can make $\bf{O}$ be transferred to $\bf{OTD}$. For more specific instructions, the proportion that $\bf{U}$ in $\bf{O}$ is transformed to the $\bf{OTD}{\bf{/}}{\kern 1pt} {\bf{O}}$ is investigated via some specific $\bf{V}$, the simulation results are shown in Table \ref{table:2}. \\

\section{CONCLUSIONS AND FUTURE WORKS}\label{sec:VI}
\indent In a general configuration, we investigate the decidability of quantum common causes and quantum causality via the statistic $C$, which has the potential to assess the existence of causality between two qubits. To this end, $C$  is extended to the real domain. It turns out that $C \in \left[ { - 1,\frac{1}{{27}}} \right]$ if two qubits are caused by quantum common causes; $C \in \left[ { - \frac{1}{{27}},1} \right]$ if two qubits are quantum causality. In addition, this paper provides an intuitive geometric interpretation of quantum common causes, quantum causality and a combination of both (see Fig. \ref{figure_2}), which can discriminate them in a different way.\\
\indent Fig. \ref{figure_2} illustrates that quantum common causes and quantum causality in Fig. \ref{figure_3} can not be discriminated via the vector-valued function $\bf{P}$. In this paper, the combination of ${\bf{P}}$ and ${\bf{P}'}$ is proposed to more effectively identify quantum common causes and quantum causality. The rationality of the combination of ${\bf{P}}$ and ${\bf{P}'}$ is well analyzed, and some simulation results are obtained. We leave a more detailed analysis on the decidability of quantum common causes and quantum causality via a combination of ${\bf{P}}$, ${\bf{P}'}$ and etc. in future works.\\

\bibliography{MainTextFile}

\end{document}